\documentclass[preprint2]{aastex}
\usepackage{spr-astr-addons}
\usepackage{graphicx}
\begin{document}
\title{Compact jets as probes for sub-parsec scale regions in AGN}

\shorttitle{Compact jets in AGN}        
\shortauthors{Lobanov}

\author{Andrei Lobanov}
\affil{Max-Planck-Institut f\"ur Radioastronomie, Auf dem H\"ugel 69, 53121 
Bonn, Germany. E-mail: alobanov@mpifr-bonn.mpg.de}

\begin{abstract}
Compact relativistic jets in active galactic nuclei offer an effective
tool for investigating the physics of nuclear regions in galaxies. The
emission properties, dynamics, and evolution of jets in AGN are
closely connected to the characteristics of the central
supermassive black hole, accretion disk and broad-line region in
active galaxies.  Recent results from studies of the nuclear regions
in several active galaxies with prominent outflows are reviewed in
this contribution.
\end{abstract}



\section{Introduction}
\label{lobanov2:sec1}

Substantial progress achieved during the past decade in studies of
active galactic nuclei has brought an increasingly wider recognition
of the ubiquity of relativistic outflows (jets) in galactic
nuclei, turning them into an effective
probe of nuclear regions in galaxies \cite{lobanov2006}. Emission
properties, dynamics, and evolution of an extragalactic jet are
intimately connected to the characteristics of the supermassive black
hole, accretion disk and broad-line region in the nucleus of the host
galaxy.  

High-resolution radio observations access directly the regions where
the jets are formed \cite{junor1999,bach2005}, and trace their evolution and
interaction with the nuclear environment (Mundell et al. 2003). These
studies, combined with optical and X-ray studies, yield arguably the
most detailed picture of the galactic
nuclei \cite{marscher2005}. Presented below is a brief summary of
recent results in this field, outlining the relation between jets,
supermassive black holes and nuclear regions in
prominent active galactic nuclei (AGN). In this respect, this review
is complementary to other recent
reviews \cite{camenzind2005,konigl2006,marscher2005} focused on
formation and propagation of extragalactic relativistic jets.

\section{Compact jets in AGN}

Jets in active galaxies are formed in the immediate vicinity of the
central black hole \cite{camenzind2005}, and they interact with every
major constituent of AGN (see Table~\ref{lobanov:tb1}.
The jets carry away a fraction of the
angular momentum and energy stored in the accretion
flow \cite{hujeirat2003} or corona (in low luminosity
AGN; Merloni \& Fabian 2002) and in the rotation of the central black
hole \cite{koide2002,komissarov2005}.

\begin{table*}[t]
\caption{Characteristic scales in the nuclear regions in active galaxies}
\label{lobanov:tb1}
\small
\begin{center}
\begin{tabular}{rccccc}\hline\hline
   & $l$ & $l_8$ & $\theta_\mathrm{Gpc}$ & $\tau_c$ & $\tau_\mathrm{orb}$ \\ 
  & [$R_\mathrm{g}$] & [pc] & [mas]& [yr] & [yr] \\ \hline
Event horizon:           &1--2          &$10^{-5}$           &$5\times 10^{-6}$    &0.0001     & 0.001 \\
Ergosphere:              &1--2          &$10^{-5}$           &$5\times 10^{-6}$    &0.0001     & 0.001 \\
Corona:                  &10$^1$--10$^2$&$10^{-4}$--$10^{-3}$&$5\times 10^{-4}$    &0.001--0.01& 0.2--0.5 \\
Accretion disk:          &10$^1$--10$^3$&$10^{-4}$--$10^{-2}$&$0.005$              &0.001--0.1 & 0.2--15 \\
{\bf Jet formation:}          &$>$10$^2$     &$>$$10^{-3}$        &$>$$5\times 10^{-4}$ &$>$0.01    & $>$0.5 \\
{\bf Jet visible in the radio:}&$>$10$^3$     &$>$$10^{-2}$        &$>$$0.005$           &$>$0.1     & $>$15 \\ 
Broad line region:       &10$^2$--10$^5$&$10^{-3}$--1        &$0.05$               &0.01--10   & 0.5--15000 \\
Molecular torus:         &$>$10$^5$     &$>$1                &$>$$0.5$             &$>$10      & $>$15000 \\
Narrow line region:      &$>$10$^6$     &$>$10               &$>$5                 &$>$100     & $>$500000 \\ \hline
\end{tabular}
\end{center}
{Column designation:}~$l$ -- dimensionless scale in units of the
gravitational radius, $G\,M/c^2$; $l_8$ -- corresponding linear scale,
for a black hole with a mass of $5\times 10^8\,M_{\odot}$;
$\theta_\mathrm{Gpc}$ -- corresponding largest angular scale at 1\,Gpc
distance; $\tau_c$ -- rest frame light crossing time;
$\tau_\mathrm{orb}$ -- rest frame orbital period, for a circular
Keplerian orbit. Adapted from~\cite{lobanov2006}
\end{table*}

\begin{figure}[t]
\includegraphics[height=0.48\textwidth,angle=-90,bb=51 62 571 686,clip=true]{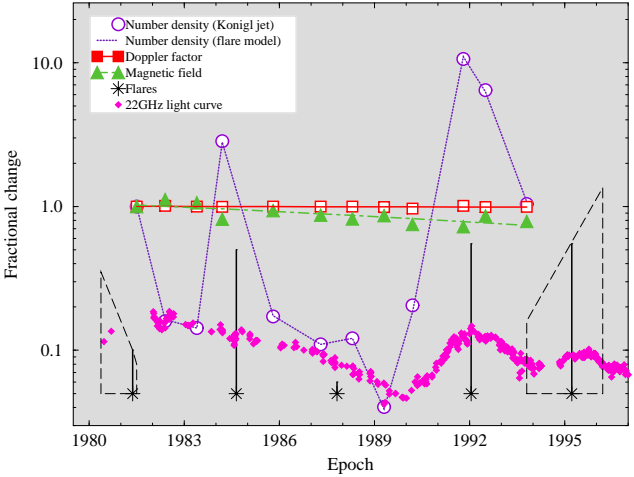}
\caption{Relative changes of the Doppler factor and
magnetic field in the ``core'' of the jet in 3C\,345, obtained by applying
the K\"onigl jet model to the measured flux density and frequency of the
synchrotron peak in the spectrum \cite{lobanov1999}. All quantities are normalized to
their respective values at the first epoch, $t_0=1981.5$. Open circles
denote the particle density required for maintaining a constant
Doppler factor. The dotted line shows the particle density, as
represented by five exponential flares. The resulting 
Doppler factor (open squares) and magnetic field (filled triangles)
are also shown, with lines representing linear fits to the respective
quantity. The total 22\,GHz flux density is plotted for
comparison, scaled down by a factor of 100.  }
\label{lobanov:fg01}
\end{figure}

The production of highly-relativistic outflows requires a large
fraction of the energy to be converted to Poynting flux in the very
central region \cite{sikora2005}.  The efficiency of this process may
depend on the spin of the central black hole \cite{meier1999}. The
collimation of such a jet requires either a large scale poloidal
magnetic field threading the disk \cite{spruit1997} or a slower and
more massive MHD outflow launched at larger disk radii by centrifugal
forces (Bogovalov \& Tsinganos 2005). 

At distances of $\sim 10^3\,R_\mathrm{g}$ ($R_\mathrm{g} = G\,M/c^2$
is the gravitational radius of a black hole), the jets become visible
in the radio regime, which makes high-resolution VLBI\footnote{Very
Long Baseline Interferometry} observations a tool of choice for
probing directly the physical conditions in AGN on such small scales.

Recent studies indicate that at distances of
$10^3$--$10^5\,R_\mathrm{g}$ ($\lesssim 1$\,pc) the jets are likely to
be dominated by pure electromagnetic processes such as Poynting flux
\cite{sikora2005} or have both MHD and electrodynamic components
\cite{meier2003}. The flowing plasma is likely to be dominated by
electron-positron pairs \cite{wardle1998,hirotani2005} although a
dynamically significant proton component cannot be completely ruled
out at the moment \cite{celotti1993}.  How far the magnetic field
dominated region extends in extragalactic jets is still a matter of
debate. There are indications that this region does not extend beyond
several parsecs. Parsec-scale flows show clear evidence for the
presence of rapidly dissipating relativistic shocks (e.g. Lobanov \&
Zensus 1999) preceeding the development of Kelvin-Helmholtz
instability \cite{lobanov2001} dominating the flow dynamics on
kiloparsec scales \cite{lobanov2003}.


{\em Ultracompact} jets observed down to sub-parsec scales typically
show strongly variable but weakly polarized emission (possibly due to
limited resolution of the observations).  Compelling evidence exists
for acceleration \cite{bach2005} and collimation \cite{junor1999} of
the flows on these scales, which is most likely driven by the magnetic
field \cite{vlahakis2004}.  The ultracompact outflows are probably
dominated by electromagnetic processes \cite{meier2003,sikora2005},
and they become visible in the radio regime (identified as compact
``cores'' of jets) at the point where the jet becomes optically thin
for radio emission \cite{lobanov1998a,lobanov1999}. At this point, the
jets do not appear to have strong shocks \cite{lobanov1998} and their
basic properties are successfully described by quasi-stationary flows
\cite{konigl1981}.  The evolution and variability of the ``core'' can
be explained by smooth changes in particle density of the flowing
plasma, associated with the nuclear flares in the central engine
(Fig.~\ref{lobanov:fg01}. Intrinsic brightness temperatures of the
ultracompact jets are estimated to be (1--$5) \times 10^{11}$\,K
\cite{lobanov2000}, implying that the energy losses are dominated by
the inverse-Compton process.

\begin{figure}[t]
\centering
\includegraphics[width=0.48\textwidth,angle=0,bb=12 12 576 480,clip=true]{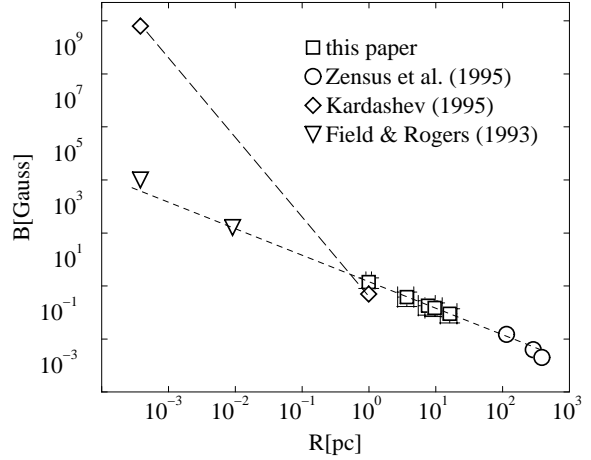}
\caption{Magnetic field distribution in the jet in
3C\,345. Squares show the magnetic field in the compact jet derived
from the frequency dependent shift of the core. Circles are the
homogeneous synchrotron model estimates of magnetic field in the
extended jet components.  Triangles show the
characteristic magnetic field values from a model of magnetized
accretion disk. Diamonds are the theoretical
estimates for the dipole magnetic field
around a supermassive rotating black hole. Reproduced from \cite{lobanov1998a}.}
\label{lobanov:fg02}
\end{figure}

Quasi-periodic variability of the radio emission from the ultracompact
jets is most likely related to instabilities and non-stationary
processes in the accretion disks around central black holes in AGN
(Igumenschev \& Abramovicz 1999; Lobanov \& Roland 2005). Alternative
explanations involve binary black hole systems in which flares are
caused by passages of the secondary through the accretion disk around
the primary \cite{ivanov1998,lehto1996}. These models however require
very tight binary systems, with orbits of the secondary lying well
within $10^3$ Schwarzschild radii of the primary (between 20 and 100
Schwarzschild radii, in the celebrated case of OJ\,287; Lehto \& Valtonen 1996), which poses inevitable problems for maintaining an
accretion disk around the primary (for massive secondaries; Lobanov
2006) or rapid alignment of the secondary with the plane of the
accretion disk (for small secondaries; Ivanov et
al. 1999). Discrepancy between the predicted and actual epoch of the
latest outburst in OJ\,287 \cite{valtonen2006} indicates further that
the observed behavior is not easy to be reproduced by a binary black
hole scenario and it is indeed more likely to result from a
quasi-periodic process in the disk, similarly to the flaring activity
observed in 3C\,345 \cite{lobanov2005b}.

\begin{figure}[t]
\centering
\includegraphics[width=0.48\textwidth,angle=0,bb=12 12 576 480,clip=true]{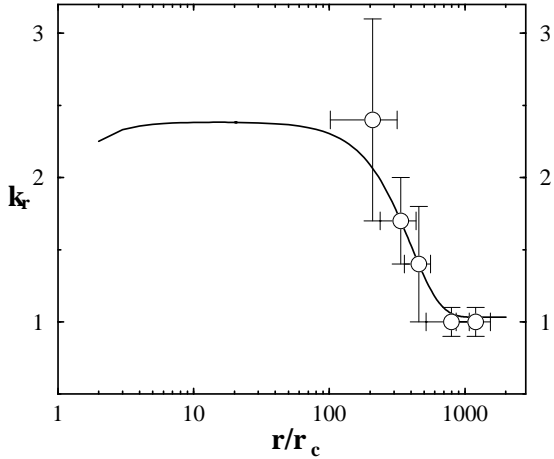}
\caption{Opacity in the jet in 3C\,309.1. Circles are the measured
values of the core shift power index $k_\mathrm{r}$ in 3C\,309.1 at
different frequencies \cite{lobanov1998}. Solid line shows changes of
$k_\mathrm{r}$ due to pressure gradients in the BLR clouds supported
by thermal pressure and maintaining a mass distribution with
spherically symmetrical gravitational potential. The cloud region
extends up to 400\,$r_\mathrm{s}$ ($r_\mathrm{s}$ refers to the
distance at which the jet becomes supersonic). The equipartition
regime is approached at the outer boundary of the cloud region, with
$k_\mathrm{r} = 1$. Significant deviations from the equipartition are
seen on smaller scales, resulting in stronger self-absorption in the
inner parts of the jet. }
\label{lobanov:fg03}
\end{figure}

\section{Jets and nuclear regions in AGN}

A number of recent studies have used the ultracompact jets as probes
of physical conditions in the central regions of AGN.  These studies
have focused on basic characteristics of relativistic
flows and atomic, physical properties of the molecular material in
circumnuclear regions of AGN, and connection between relativistic
outflows and accretion disks and broad-line emitting regions.

Synchrotron self-absorption and external absorption in the
ultracompact jets (VLBI ``cores'') can be used effectively for
determining the properties of the flow itself and its environment
\cite{lobanov1998a}.  Absolute position of the core, $r_\mathrm{c}$,
varies with the observing frequency, $\nu$, so that $r_\mathrm{c}
\propto \nu^{-1/k_\mathrm{r}}$, \cite{konigl1981}. If the core is
self-absorbed and in equipartition, the power index $k_\mathrm{r}=1$
\cite{blandford1979}.  Changes of the core position measured between
three or more frequencies can be used for determining the value of
$k_\mathrm{r}$ and estimating the strength of the magnetic field,
$B_\mathrm{core}$, in the nuclear region and the offset,
$R_\mathrm{core}$, of the observed core positions from the true base
of the jet (see Fig.~\ref{lobanov:fg02}).  The combination of
$B_\mathrm{core}$ and $R_\mathrm{core}$ gives an estimate for the mass
of the central black hole $M_\mathrm{bh} \approx 7\times
10^9\,M_\odot\, (B_\mathrm{core}/\mathrm{G})^{1/2}
(R_\mathrm{core}/\mathrm{pc})^{3/2}$.

Core shift measurements provide estimates of the total (kinetic +
magnetic field) power, the synchrotron luminosity, and the maximum
brightness temperature, $T_\mathrm{b,max}$ in the jets can be made
\cite{lobanov1998}. The ratio of particle energy to magnetic field
energy can also be estimated, from the derived $T_\mathrm{b,max}$.
This enables testing the original K\"onigl model \cite{konigl1981} and
several of its later modifications (e.g., Hutter \& Mufson 1986; Bloom
\& Marscher 1996). The known distance from the nucleus to the jet
origin will also enable constraining the self-similar jet model
\cite{marscher1995} and the particle-cascade model
\cite{blandford1995}.

Recent studies of free-free absorption in AGN indicate the presence of
dense, ionized circumnuclear material with $T_\mathrm{e} \approx
10^4$\,K distributed within a fraction of a parsec of the central
nucleus \cite{lobanov1998a,walker2000}.  Properties of the
circumnuclear material can also be studied using the variability of
the power index $k_\mathrm{r}$ with frequency. This variability
results from pressure and density gradients or absorption in the
surrounding medium most likely associated with the broad-line region
(BLR). Changes of $k_\mathrm{r}$ with frequency, if measured with
required precision, can be used to estimate the size, particle density
and temperature of the absorbing material surrounding the jets (see
Figure~\ref{lobanov:fg03}). Estimates of the black hole mass and size of
the BLR obtained from the core shift measurements can be compared with
the respective estimates obtained from the reverberation mapping and
applications of the $M_\mathrm{bh}$--$\sigma_\star$ relation.

\subsection{Atomic and molecular absorption}

Opacity and absorption in the nuclear regions of AGN have been probed
effectively using the non-thermal continuum emission as a background
source. Absorption due to several atomic and molecular species (most
notably due to H\i, CO, OH, and HCO$^+$) has been detected in a number
of extragalactic objects. OH absorption has been used to probe the
conditions in warm neutral gas \cite{goicoechea2004,kloeckner2005},
and CO and H\i\ absorption were used to study the molecular tori
\cite{conway1999,pedlar2004} at a linear resolution often smaller than
a parsec \cite{mundell2003}. These observations have revealed the
presence of neutral gas in a molecular torus in NGC\,4151 and in a
rotating outflow surrounding the relativistic jet in 1946$+$708
\cite{peck2001}.

\subsection{Jet-disk and jet-BLR connections}

\begin{figure}[t]
\centering
\includegraphics[width=0.48\textwidth,angle=0,bb=0 0 821 490,clip=true]{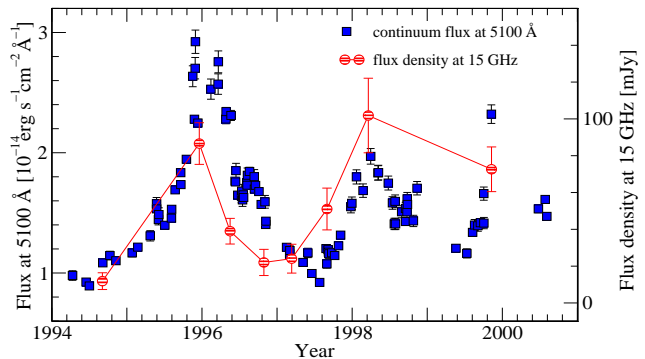}
\caption{Relativistic jet produces the bulk of non-thermal, ionising
continuum driving the broad-line emission in 3C\,390.3
\cite{arshakian2006}.  This is manifested by a correlation between the
optical continuum flux at 5100\AA\ (blue
squares) and the radio flux density at 15 GHz from a
stationary component S1 (red circles) located at about 0.4 pc distance
from the base of the jet.}
\label{lobanov:fg04}
\end{figure}

Connection between accretion disks and relativistic outflows
\cite{hujeirat2003} has been explored, using correlations between
variability of X-ray emission produced in the inner regions of
accretion disks and ejections of relativistic plasma into the flow
\cite{marscher2002}. The jets can also play a role in the generation
of broad emission lines in AGN (Fig.~\ref{lobanov:fg04}). The beamed
continuum emission from relativistic jet plasma can illuminate atomic
material moving in a sub-relativistic outflow from the nucleus,
producing broad line emission in a conically shaped region located at
a significant distance above the accretion disk
\cite{arshakian2006}. Magnetically confined outflows can also contain
information about the dynamic evolution of the central engine, for
instance that of a binary black hole system \cite{lobanov2005b}. This
approach can be used for explaining, within a single framework, the
observed optical variability and kinematics and flux density changes
of superluminal features embedded in radio jets.

\section{Conclusion}

Extragalactic jets are an excellent laboratory for studying physics of
relativistic outflows and probing conditions in the central regions of
active galaxies. Recent studies of extragalactic jets show that they
are formed in the immediate vicinity of central black holes in
galaxies and carry away a substantial fraction of the angular momentum
and energy stored in the accretion flow and rotation of the black
hole. The jets are most likely collimated and accelerated by
electromagnetic fields. 
Convincing
observational evidence exists connecting ejections of material into
the flow with instabilities in the inner accretion disks. In
radio-loud objects, continuum emission from the jets may also drive
broad emission lines generated in sub-relativistic outflows
surrounding the jets.  Magnetically confined outflows may preserve
information about the dynamics state of the central region, allowing
detailed investigations of jet precession and binary black hole
evolution to be made. This makes studies of extragalactic jets a
powerful tool for addressing the general questions of physics and
evolution of nuclear activity in galaxies.

\end{document}